\documentclass{webofc}
\usepackage[varg]{txfonts}   
\usepackage{subcaption}
\usepackage{romannum}
\usepackage[colorlinks=true, linkcolor=blue, urlcolor=blue, citecolor=blue, anchorcolor=blue]{hyperref}

\begin{document}

\title{The use of Convolutional Neural Networks for signal-background classification in Particle Physics experiments}

\author{\firstname{Venkitesh} \lastname{Ayyar}\inst{1}\fnsep\thanks{\email{vpa@lbl.gov}} \and
	\firstname{Wahid} \lastname{Bhimji}\inst{1} \and
        \firstname{Lisa} \lastname{Gerhardt}\inst{1} \and
	\firstname{Sally} \lastname{Robertson}\inst{2} \and
	\firstname{Zahra} \lastname{Ronaghi}\inst{3}    
}

\institute{Lawrence Berkeley National Laboratory, 1 Cyclotron Rd, Berkeley, California 94720, USA. 
\and
           University of California, 101 Sproul Hall, Berkeley, California 94720, USA.
\and
           NVIDIA, 2788 San Tomas Expressway, Santa Clara, California 95051, USA.}

\abstract{%
The success of Convolutional Neural Networks (CNNs) in image classification has prompted efforts to study their use for classifying image data obtained in Particle Physics experiments. Here, we discuss our efforts to apply CNNs to 2D and 3D image data from particle physics experiments to classify signal from background. 

In this work we present an extensive convolutional neural architecture search, achieving high accuracy for signal/background discrimination for a HEP classification use-case based on simulated data from the Ice Cube neutrino observatory and an ATLAS-like detector. We demonstrate among other things that we can achieve the same accuracy as complex ResNet architectures with CNNs with less parameters, and present comparisons of computational requirements, training and inference times.
}
\maketitle
\section{Introduction} \label{intro}
Particle physics experiments have been incredibly successful in improving our understanding of nature. 
An important aim of many of these experiments is to search for elusive particles with interesting properties. Evidence for these particles comes from rare events recorded in the detectors of these experiments, termed {\it signal}.
These are accompanied by large number of other events from known and well-tested particle interactions, called { \it background}. 
To extract the interesting physics, it is essential to filter out the background from the signal. This process of distinguishing between the two is known as {\it Signal-background} classification and lies at the heart of experimental particle physics. Currently this is achieved by applying selections on derived high-level physics variables. 

Given the success of deep learning methods in classifying real-world images, deep learning methods have been applied to the problem of signal-background classification (\cite{ice-1},~\cite{atlas2} and references therein). In this work, we describe the use of Convolutional Neural Networks (CNNs) to classify signal from background for two use cases: a simulation dataset for the IceCube experiment and one for the ATLAS experiment.
 
\section{Convolutional Neural Networks}
Convolutional Neural Networks (CNNs) are a class of neural networks specialized for {\it image classification}.
They are designed to capture features of images at different scales.

\begin{figure}[h]
\centering
\includegraphics[width=.8\linewidth]{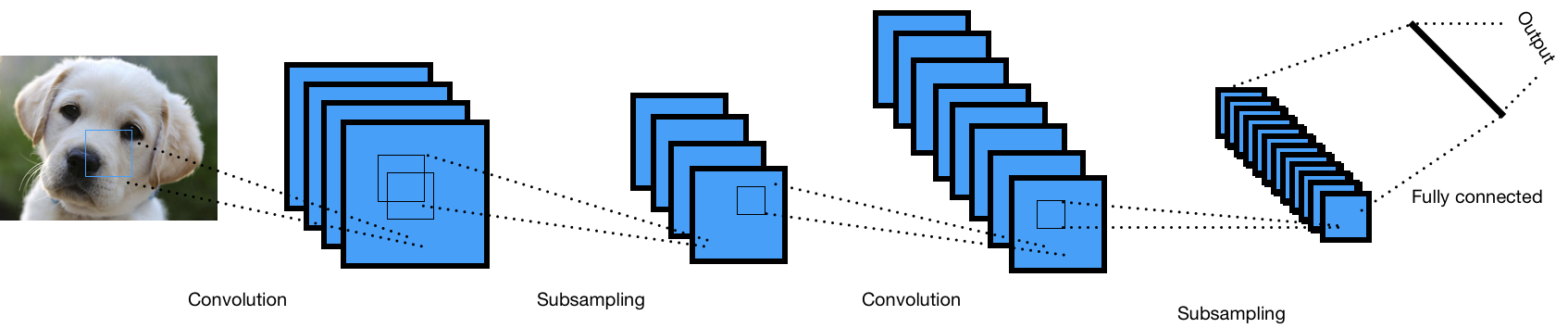}
\caption{The general structure of a CNN with {\it convolutional}, {\it subsampling} and {\it fully-connected} layers stacked together.}
\label{fig-cnn}       
\end{figure}

Fig.~\ref{fig-cnn} shows the basic structure of a CNN. It typically starts with a {\it Convolutional} layer: scanning through the input image in small blocks, it performs convolution operations to create feature maps. The dimensions of the resulting layers are large and they are reduced by the subsequent {\it Subsampling} layers. Stacking many such blocks sequentially can enable the networks to learn a variety of features at different scales. Eventually the neurons are combined together using one or more {\it Fully connected} layers. The output of final layer predicts the image class.
CNNs have been very successful in classifying 2D images ~\cite{cnn-mnist,cnn-lenet}.

\section{Dataset 1: IceCube}
\subsection{The IceCube experiment}
\begin{figure}
\centering
\begin{subfigure}{.5\textwidth}
  \centering
  \includegraphics[width=1.0\linewidth]{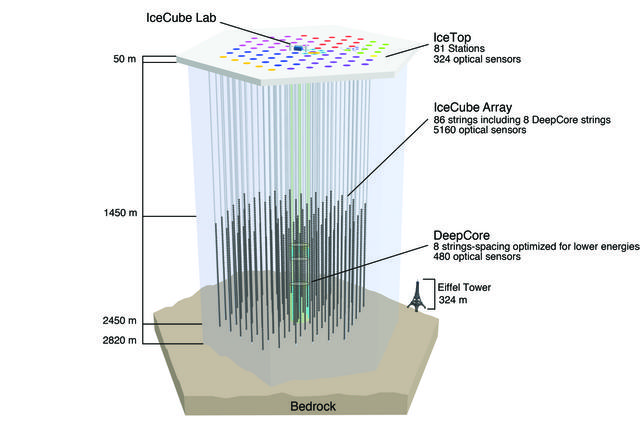}
  \caption{}
  \label{fig:ice1}
\end{subfigure}%
\hfill
\begin{subfigure}{.50\textwidth}
  \centering
  \includegraphics[width=1.0\linewidth]{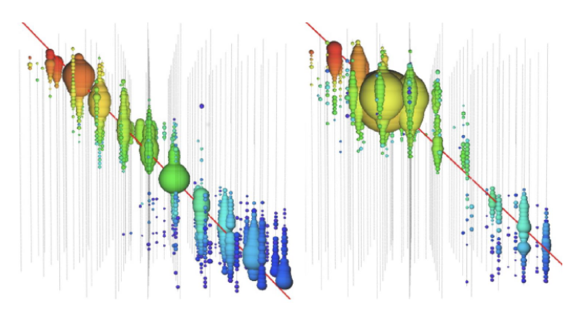}
  \caption{}
  \label{fig:ice2}
\end{subfigure}
\caption{ \small Fig (a) shows the setup of the IceCube Observatory. There are approximately 5000 Digital Optical modules (DOMs) placed within the Antarctic ice. Fig (b) shows the pattern of light deposition for signal and background. The colored bubbles indicate the relative time of arrival of light, with red being earliest and blue being the latest. The size of the bubbles is proportional to the number of photons. To the left, the pattern is more uniformly distributed, which is indicative of a background event produced by atmospheric muons. 
The pattern in the right figure has a large stochastic energy deposition which is indicative of a signal due to a muon produced from an astrophysical neutrino.}
\label{fig:Icecube}
\end{figure}

IceCube is a neutrino observatory located at the South Pole, looking for high energy (>100GeV) astrophysical neutrinos~\cite{ice-2,ice-3}. Interactions of these high energy neutrinos with nuclei produce secondary charged particles. These emit Cherenkov light and are detected using an array of Digital Optical Modules (DOMs) placed below the ice. The experimental setup is depicted in Fig.~\ref{fig:ice1}. 

While different particles can contribute to the Cherenkov light seen by the DOMs, this dataset only considers the contribution of {\it muons}. The Cherenkov radiation from muons produced by astrophysical neutrinos forms the {\it signal} for this dataset. The {\it background} consists of the light contribution from other atmospheric muons. 
This dataset used the high energy down-going region of IceCube detection, a region not used in most analysis due to the high background component.  The physical way to distinguish the signal and background in this region is using the stochasticity of energy deposition. Signal muons obtained from reactions of astrophysical neutrinos are single muons and hence lose energy stochastically. This results in uneven light emission along the track. The atmospheric muons  typically consist of hundreds of muons and hence their light emission averages out, resulting in a more even distribution. This is shown in Fig.~\ref{fig:ice2}.

\subsection{Data set}
The input data samples comprise of {\it events}, each consisting of the timings and the total charge deposited on each DOM. Each raw input image has the dimensions 86 x 60 laid out on a hexagonal grid, and is mapped onto an orthogonal grid to create an image of size 10 x 20 x 60. More details about the dataset can be found in ~\cite{ice-1}.
This study did not use information from the Deep Core DOMs, which are specialized DOMs placed near  the center of the IceCube grid.

In ~\cite{ice-1}, a few of the authors in this paper had explored the potential of Graph Neural networks (GNNs) and CNNs in performing signal-background classification. Their results showed that both GNNs and CNNs  performed better than the physics benchmarks, with the GNNs achieving better performance than the CNNs used in the paper (ResNet).
In this work, we perform a more through exploration of CNN architectures. Another aspect that differentiates this work is the exclusion of a set of events called High Energy Starting Events (HESE)~\cite{ice-4}, which are neutrino events where the interaction starts inside the detector, and can be identified in a pre-selection stage using existing IceCube analyses. 

For training and validation, we used 130989 samples, with a validation ratio of 33\% and signal to background ratio of 16.2\%. The test dataset had 737715 samples with a signal to background ratio of 1.92\%.

\subsection{Analysis and Results} \label{ice:analysis}
The study in~\cite{ice-1} utilized the {\it ResNet} CNN (similar to the ones in~\cite{resnet}), with very high number of parameters. In this analysis, we did a more thorough architecture search, specifically looking for compact, layered 3D CNNs with lesser number of parameters. Scanning CNN network architectures using combinations of convolution, pooling and dropout layers, we identified a few networks that achieved better performance than the previous ResNet model.


\begin{figure}
\centering
\begin{subfigure}{.50\textwidth}
  \centering
  \includegraphics[width=1.0\linewidth]{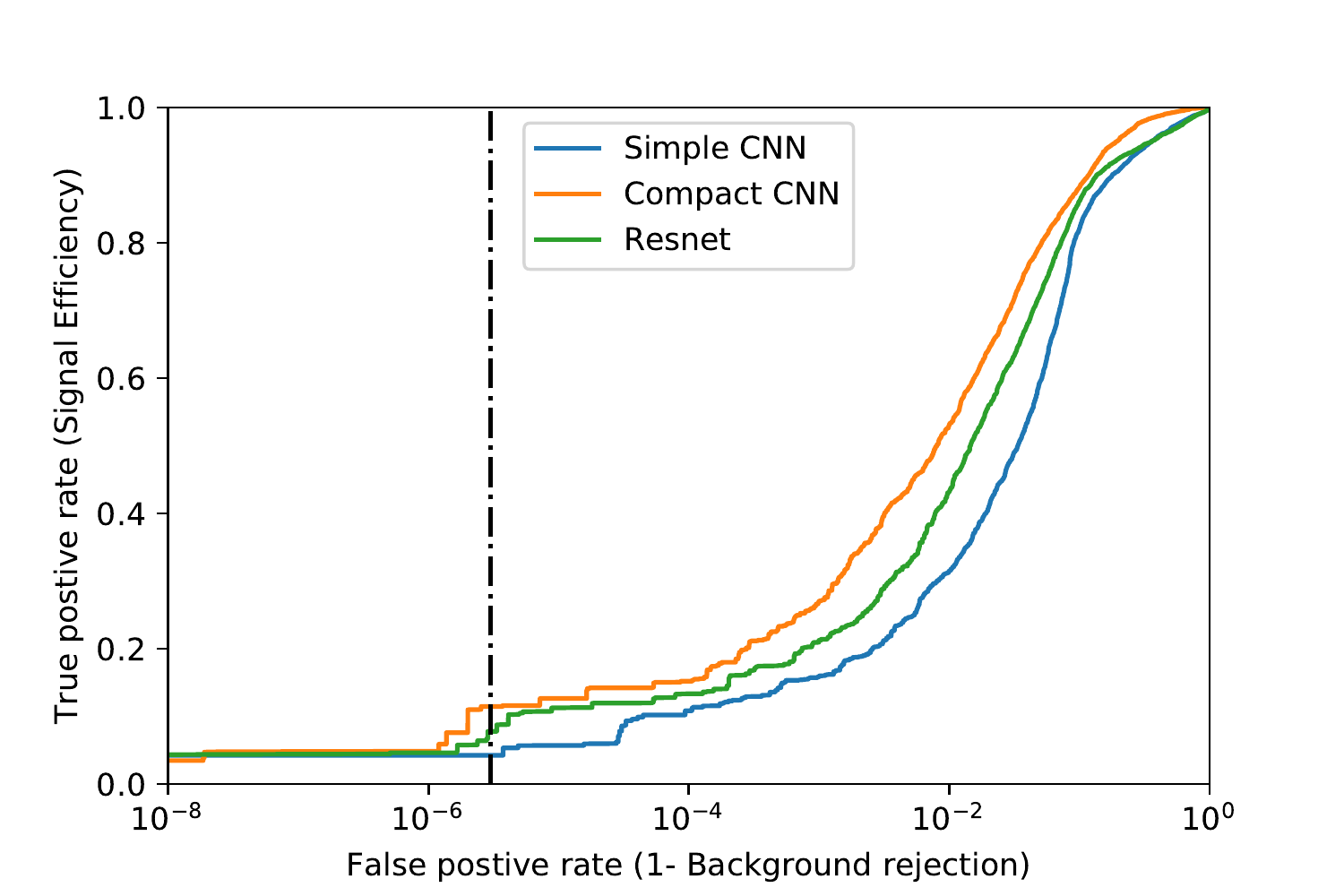}
  \caption{ \small Roc curves for 3 chosen models. The X axis denotes the false positive rate(fpr) and the Y axis denotes the True positive rate(tpr). The dotted line denotes an fpr = $ 3 \mbox {x} 10^ {-6} $. The model {\it Compact CNN} from this work achieves best performance everywhere. We also compare these with a simpler model with very few parameters termed {\it Simple CNN}}
  \label{fig:ice_roc}
\end{subfigure}%
\hfill
\begin{subfigure}{.45\textwidth}
  \centering
  \includegraphics[width=1.0\linewidth]{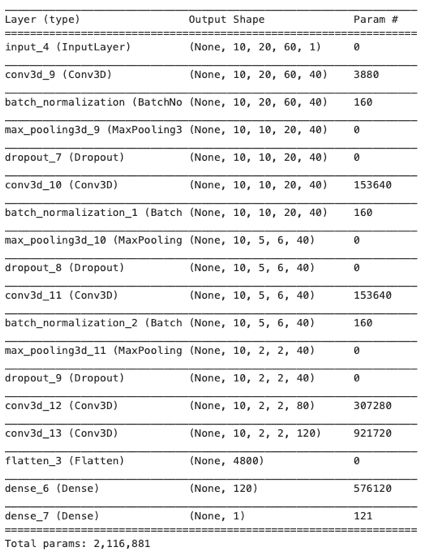}
  \caption{Structure of the best performing model. It has $ \sim 2 $ million parameters.}
  \label{fig:ice_bestmodel}
\end{subfigure}
\caption{}
\label{fig:icecube_results}
\end{figure}

\begin{table}
\centering
\caption{Comparison of selected trained models. {\it tpr} and {\it fpr} stand for true positive rate and false positive rate respectively, while {\it AUC} stands for area under the  ROC curve. {\it Simple CNN} is a layered CNN with very few number of parameters, {\it Compact CNN} is the best performing model and ResNet is the model used in the previous paper ~\cite{ice-1}. The training times are obtained by running on a Titan X (Pascal) GPU.}
\label{tab-ice}  
\begin{tabular}{ | l| | r | r | r | r|}
\hline
Model type & \# of parameters & Training time (s / epoch ) & AUC score & Tpr at fpr=$ 3 \mbox{x} 10^{-6} $ \\
\hline
\hline
Simple CNN & 14,789 & 23 & 0.922 & 0.04\\
\hline
Compact CNN & 2,116,881 & 120 & 0.960 & 0.116 \\
\hline
ResNet & 26,486, 126 & 115 & 0.935 & 0.078 \\
\hline
\end{tabular}
\end{table}

A good way to assess the performance of models is by looking at the Receiver Operating Characteristic (ROC) curves. Fig. ~\ref{fig:ice_roc} shows the ROC curves for three models: ResNet, the best CNN in this work termed {\it Compact CNN} and a CNN with very few parameters termed {\it Simple CNN} and Table~\ref{tab-ice} gives a comparison of these models. 
Comparing the true positive rate (tpr) values for the three models at a false positive rate (fpr) value of $ 3 \mbox {x} 10^ {-6} $ (which was the value used for comparison with the physics cuts in \cite{ice-1}), it is clear that the {\it Compact CNN} performs better than ResNet, while having only a tenth of the parameters. 
Its structure is shown in Fig. ~\ref{fig:ice_bestmodel}.
Comparing the training times per epoch for the models in Table~\ref{tab-ice}, it can be seen that the training time for the {\it Compact CNN} model is slightly higher than ResNet. Given that it has only a tenth of the parameters, this is a bit surprising. It is possible that ResNet might be faster due to the connectivity between its different layers that is absent for our models.
Nevertheless, we have developed a CNN with better classification performance than ResNet, while reducing the parameters by an order of magnitude.

\section{Dataset 2: ATLAS }
\subsection{ATLAS experiment}
The ATLAS experiment is one of the major experiments at the Large Hadron Collider (LHC) at CERN, Switzerland. It was one of the two LHC experiments involved in the discovery of the Higgs boson in 2012. Among other things, one of the main goals of the LHC is to look for evidence for Physics beyond the Standard Model of particle physics, such as {\it Supersymmetry}. In this work, we use a dataset that is an input for analyses searching for new massive supersymmetric (RPV-Susy) particles in multi-jet final states~\cite{atlas1}. The goal is to use CNNs directly on low-level detector data from the entire calorimeter, without reconstruction of jets or tuning of analysis variables. 
\subsection{Data set}
The dataset consists of simulated data obtained using the {\it Pythia} event generator~\cite{pythia} interfaced to the {\it Delphes fast detector simulation}~\cite{delphes,fastjet}. The Signal events are the RPV-Susy events, while the background is QCD. The images are 2D, with dimensions 64 x 64, with each image pixel representing the energy deposited in the calorimeter.
For training and validation and testing, we used 412416, 137471 and 137471 samples respectively, with a signal to background ratio of about 43\%.

\subsection{Analysis and Results}
In ~\cite{atlas2}, the potential of CNNs for signal-background classification was explored with this dataset. A few simple CNNs with large number of parameters were studied and they were found to achieve better performance than the physics benchmarks. The best model was found to have around 34 million parameters. The main aim of this work was to perform an architecture search to identify CNNs with better performance and fewer parameters.

\begin{figure}
\centering
\begin{subfigure}{.50\textwidth}
  \centering
  \includegraphics[width=1.0\linewidth]{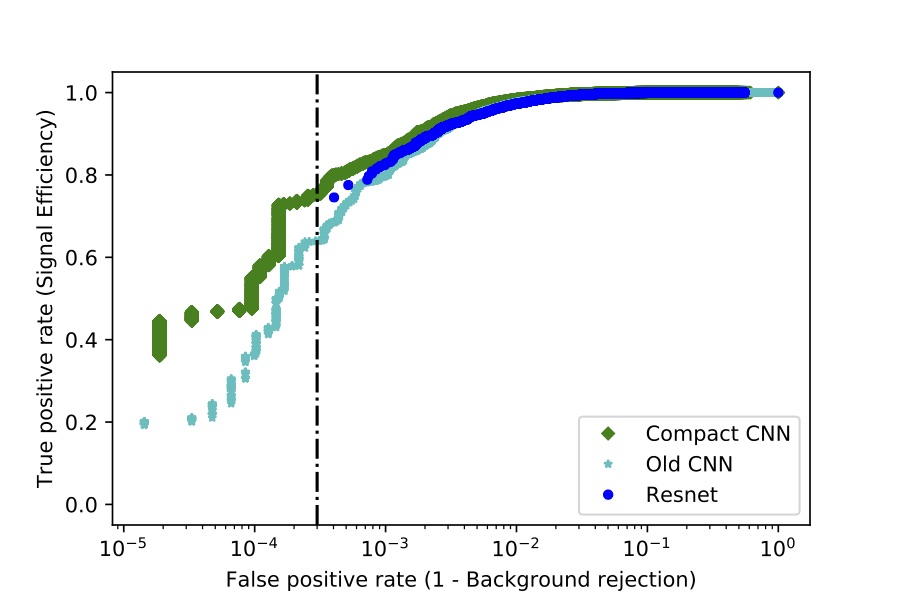}
  \caption{ \small Roc curves for 3 chosen models. The X axis denotes the false positive rate(fpr) and the Y axis denotes the True positive rate(tpr). The dotted line denotes an fpr = $ 3 \mbox {x} 10^ {-6} $. The Compact CNN achieves best performance everywhere, excelling especially in the low fpr region to the left.}
  \label{fig:atlas1}
\end{subfigure}%
\hfill
\begin{subfigure}{.45\textwidth}
  \centering
  \includegraphics[width=1.0\linewidth]{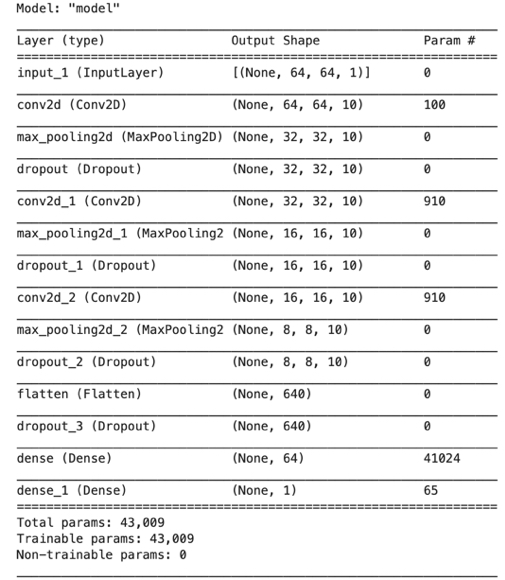}
  \caption{Structure of the best performing model. It has $ \sim 43,000 $ parameters.}
  \label{fig:atlas2}
\end{subfigure}
\caption{Performance of compact layered CNNs.}
\label{fig:atlas}
\end{figure}

As in the previous case, we explored network architectures using a combination of convolution, pooling and dropout layers. Table \ref{tab-atlas2} and Fig.~\ref{fig:atlas1} show a comparison of the best performing model termed {\it Compact CNN} with the model used in the previous work and ResNet. From the ROC curves for these 3 models in Fig.~\ref{fig:atlas1}, it is clear that Compact CNN performs better than the other two models. It is also much simpler in structure (almost 1/800th the number of parameters) compared to the other two models, while having a significantly lower training time per epoch as seen in Table \ref{tab-atlas2}. The structural details of the Compact CNN model is given in Fig.~\ref{fig:atlas2}.
Thus, we have developed a substantially more compact network, while reducing the number of parameters by almost 3 orders of magnitude.

\begin{table}
\centering
\caption{Comparison of selected trained models. {\it tpr} and {\it fpr} stand for true positive rate and false positive rate respectively. {\it Old CNN} denotes the CNN used in the previous work, {\it Compact CNN} is the best performing CNN in this work. We also compare these with a ResNet model. The training times are obtained by running on a Titan X (Pascal) GPU.}
\label{tab-atlas2}  
\begin{tabular}{ | l |r |r |r||}
\hline

 Model type & \# of parameters & Training time (s / epoch)  & tpr at fpr=$ 3 \mbox{x} 10^{-3} $\\
\hline
\hline
Old CNN & 34,515,201 & 294 & 0.641 \\
\hline
ResNet & 23,597,826 & 515 & - \\
\hline 
Compact CNN & 43,009 & 40 & 0.746 \\
\hline
\end{tabular}
\end{table}

\section{Summary}
We have demonstrated the effectiveness of compact, layered Convolutional Neural Networks in classifying signal from background for simulated datasets from two different particle physics experiments. In both cases, performing an architecture search, we have identified compact neural networks that achieve better performance than previous studies with substantially fewer parameters.
Given this success of CNNs in signal-background classification, there is potential for their applicability to classification problems in other fields beyond particle physics.

\section{Computational details}
All computations were performed at NERSC. The CNNs were implemented in {\it keras}~\cite{keras}.
During the course of this work, we developed a package for training of general stacked CNN models with visualization tools\footnote[1]{https://github.com/vpayyar/layered\_CNNs\_for\_ATLAS\_data}. These should be of use for general purpose CNN training and visualization
\section{Acknowledgments}
We would like to thank the IceCube collaboration for providing access to the simulation dataset. We would also like to thank Spencer Klein and Nick Choma for useful discussions about the IceCube part of this work. 
This research used resources of the National Energy Research Scientific Computing Center (NERSC), a U.S. Department of Energy Office of Science User Facility operated under Contract No. DE-AC02-05CH11231.
V.A.'s work was supported by the Computational Center for Excellence, a Computational HEP program in the Department of Energy's Science Office of High Energy Physics (Grant \#KA2401022).
S.R's work was supported by the National Science Foundation under grant number PHY-1307472.

\end{document}